\documentclass[twocolumn]{svjour3}         
\smartqed  
\usepackage{graphicx}
\begin{document}

\title{Evolution of Non-Equilibrium Profile in Adsorbate Layer
under Compressive Strain
}


\author{Enzo Granato        \and
        S.C. Ying 
}


\institute{Enzo Granato \at
               Laborat\'orio Associado de Sensores e Materiais,
Instituto Nacional de Pesquisas Espaciais, \\
12227-010 S\~{a}o Jos\'e dos Campos, S\~ao Paulo, Brazil \\
           \and
          S.C. Ying \at
               Department of Physics,
Brown University, \\
Providence, Rhode Island 02912, USA \\
\email{seechen.y@gmail.com} }

\date{Received: date / Accepted: date}

\maketitle

\begin{abstract}
We investigate the time evolution of an initial step profile
separating a bare substrate region from the rest of the
compressively strained adsorbate layer near a commensurate to
incommensurate transition. The rate of profile evolution as a
function of the mismatch, coverage and the strength of the substrate
potential are determined by Brownian molecular dynamics simulations.
We find that the results are qualitatively similar to those observed
for the Pb/Si(111) system. The anomalously fast time evolution and
sharpness of the non-equilibrium profile can be understood through
the domain wall creation at the boundary and its subsequent
diffusion into the interior of the adsorbate layer.

\end{abstract}

\section{Introduction}
\label{intro} The system of an adsorbate layer with a lattice
mismatch to the underlying substrate has been extensively studied
both from the experimental and theoretical side. In the monolayer
regime, it provides a realization of many two dimensional phases
\cite{Pokrov} and has also been used as a simple model for studying
sliding friction phenomena on surfaces
\cite{Persson93,Persson,Gy00,Gy04}. In particular, the rich
phenomena of commensurate to incommensurate phase transition depends
on the competition between the adsorption energy and the strain
energy from the lattice mismatch. Recently, Tringides {\it et al.}
\cite{Man,conf} have done an extensive study of the Pb/Si(111)
system and found the "Devil's Stair Case"  within a narrow range of
coverage as predicted by the theory. For a larger coverage, they
have further found that the adsorbate layer exhibited anomalous mass
transport through the time evolution of step profile created by
desorbing the adatoms in a small region. This anomalous transport is
characterized by a sharp non-diffusive profile at all times as well
as a critical coverage below which the rate of the profile evolution
slows down exponentially with decreasing coverage. To date, there is
still no detailed understanding of the microscopic origin of this
anomalous mass transport. Very recently, Huang {\it et al.}
\cite{Wang} have used molecular dynamics simulations in a
generalized one-dimensional Frenkel-Kontorova model to demonstrate
that collective diffusion of Pb atoms in the wetting layer on
Si(111) via domain wall motion for coverage beyond monolayer can
lead to fast mass transport. However, the microscopic origin of a
critical coverage for the profile evolution is still not understood.
In this paper, we present a study of a generic two dimensional
adsorbate system \cite{Persson,Gy00,Gy04} with a well understood
transition from a commensurate  $c(2\times 2)$ phase to an
incommensurate phase.  A non-equilibrium profile is  created in the
adsorbate layer and  the time evolution of the profile is then
studied numerically via molecular dynamics simulation.  In
particular, we focus on the rate of profile evolution as a function
of the lattice mismatch, the coverage and the strength of the
substrate potential. We find that the results are qualitatively
similar to those observed for the Pb/Si(111) system. The main
results can be understood via microscopic mechanisms that are
generic to an adsorbate layer under a compressive strain near a
commensurate-incommensurate transition. In particular, the time
evolution of the non-equilibrium profile is governed by the rate of
domain wall creation at the boundary and its subsequent diffusion
into the interior of the adsorbate layer.

\section{Model}
We consider a two-dimensional (2D) adsorbate layer with particle
interactions modeled by the Lennard-Jones (LJ) pair potential,

\begin{equation}
v(r) = \epsilon_0 [(r_o/r)^{12} - 2 (r_o/r)^6 ],
\end{equation}
with $2$ parameters: minimum energy distance $r_o$ and amplitude
$\epsilon_o$. The substrate is represented by a rigid sinusoidal
periodic potential,

\begin{equation}
V({x,y}) = V_0 \ [2-\cos(2\pi x/a) - \cos(2 \pi y/a)] ,
\end{equation}
with amplitude $V_o$ and minima on the sites of a square lattice
with lattice spacing $a$. This model has been used previously in the
study of nonlinear sliding friction of adsorbed overlayers
\cite{Persson93,Persson,Gy00,Gy04}. The system consists of
$5000-10000$ adatoms in an area of size $L \times L$. Periodic
boundary conditions are applied. Throughout this paper, we will
choose the amplitude of the substrate potential and the substrate
lattice spacing as the scales for the energy and length and all
quantities will be expressed in dimensionless form. In Fig. 1, we
show the ground state energy of the adsorbate system at half
coverage, $\theta =1/2$, for different values of the parameter $r_o$
in the LJ potential. For $r_o <1.56$, the adsorbate layer is in a
commensurate  $c(2\times 2)$ phase with interparticle spacing of
$\sqrt 2 $. The ground state energy is a minimum at $r_0=1.45$.
Changing the parameter $r_o$ amounts to changing the lattice
mismatch or changing the compressive strain in the commensurate
phase. For  $r_0>1.45$, the layer is compressed relative to the
minimum energy configuration but still remains in the commensurate
c(2x2) phase as seen in the constancy of the amplitude of  the
structure factor at wavevector $Q=(\pi,\pi)$. For $r_0>1.56$, the
value of the structure factor at wavevector $Q=(\pi,\pi)$ starts to
decrease, indicating the entrance into an incommensurate phase.
\begin{figure}[btp]
\includegraphics[bb= -0.5cm -0cm  9cm   10cm, width=7.5cm]{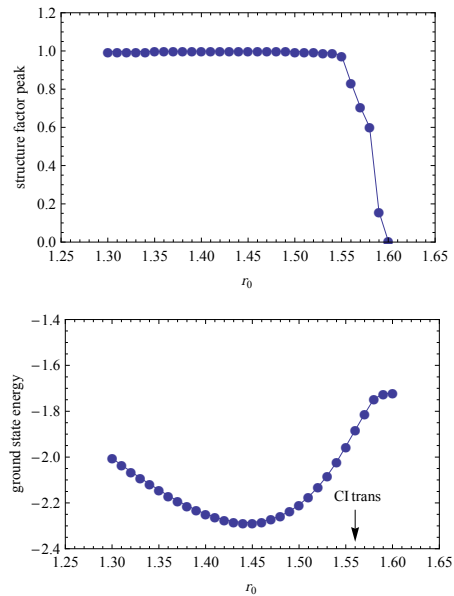}
\caption{Structure factor peak at wavevector  $Q=(\pi,\pi)$ and
ground state energy as a function of the minimum energy separation
$r_o$ in the LJ pair potential. For coverage $\theta=0.5$.}
\end{figure}

In the next two sections, we will present results when the adsorbate
atoms are removed from a circular region with radius $R << L $  to
mimic the experimental creation of a step profile \cite{Man,conf}
via localized desorption. We will then follow the temporal evolution
of this non-equilibrium profile to explore the dynamics of this
system. These studies will be performed by varying either the
mismatch through the variation of $r_0$, the coverage or the
strength of the substrate potential $V_0$. The calculations were
performed using Brownian molecular dynamics (MD) simulations
\cite{Allen} with the time variable discretized in units of $\delta
t=0.002 - 0.01 \tau_{md}$, where $\tau_{md}=(ma^2/V_o)^{1/2}$.

\section{Numerical results and discussion}

\subsection{Non-equilibrium profile evolution as a function of  lattice mismatch}

For studying the temporal evolution of the non-equilibrium profile,
the adsorbate layer is first equilibrated at finite temperature.
Then, adatoms are removed from a circular region in the center of
the system. The radius of the hole is small, corresponding to a
removal of only 150 adatoms for the system of 5000 particles. Thus,
the density of the region outside does not change significantly as
the non-equilibrium profile approaches equilibrium. Configurations
are stored at regular time intervals.

\begin{figure}[btp]
\includegraphics[bb= -0.5cm -0cm  8cm   9cm, width=7.5cm]{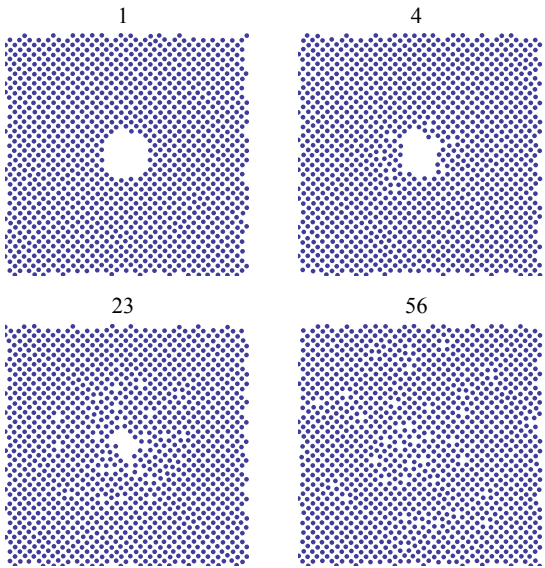}
\caption{Snapshots for the evolution of the non-equilibrium profile
for increasing times. For $r_o = 1.53$, $L=50$, temperature $T=0.4$
and coverage $\theta=0.5$. The labels correspond to times in units
of $10^4$ MD time steps.}
\end{figure}

In Fig. 2, we show snapshots for the evolution of the
non-equilibrium profile for increasing times. The parameters are
$r_o = 1.53$, $L=50$, temperature $T=0.4$  and coverage $0.5$. For
this set of parameters, the initial equilibrium phase before the
creation of the non-equilibrium profile region is the commensurate
$c(2\times 2)$ phase with interparticle spacing $\sqrt 2 $. We note
that the profile remains sharp initially as distinct from the
typical diffuse profile resulting from pure diffusion dynamics. This
is qualitatively similar to the experimental observation for
Pb/Si(111), although the small radius of the hole region prevents us
from determining a conclusive behavior for the longtime region where
the noise is significant. To study the time evolution of the profile
as a function of the lattice mismatch, simulations studies were also
done for different values of $r_0$ at half coverage. The time $\tau$
(in MD steps) that it takes to completely  refill the circular
region  and return to the equilibrium is obtained and the average
refilling velocity is given by $1/\tau$. The number of particles
inside the circular region and the radius difference $(R_o - R(t))$,
where $R_o$ is the initial radius of the hole, are both increasing
functions of time. Figs 3 show the behavior of these quantities for
$r_o=1.51$ and $ 1.56$. Note that the time scale for the two
different values of $r_0$ differ by more than an order of magnitude
showing that the refilling dynamics speed up rapidly beyond
$r_o=1.51$.
\begin{figure}[btp]
\includegraphics[bb= 0.5cm -0cm  10cm   7cm, width=7.5cm]{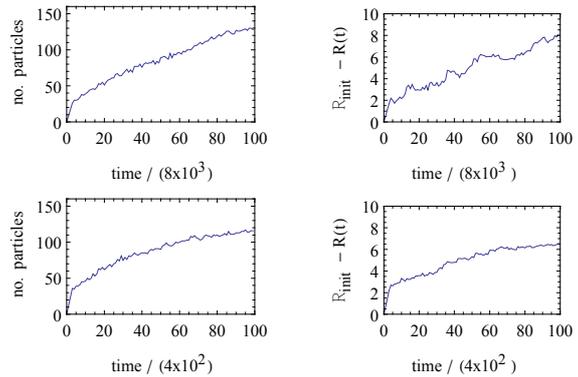}
\caption{Time dependence of the number of particles inside the hole
region ($R < R_o$) and the radius difference $R_o -R(t)$, for
$T=0.4$, $L=100$ and coverage $\theta=0.5$. For $r_o=1.51$ (top) and
$r_o=1.56$ (bottom) }
\end{figure}

In Fig. 4, we plot the average refilling velocity vs $r_0$. There is
a critical value of $r_0 \approx 1.50 $ below which the refilling
velocity ($1/\tau$) is too small to measure in our simulation study.
This value is smaller than the critical value of  $r_0 \approx
1.55$, corresponding to the bulk commensurate-incommensurate
transition at the temperature $T=0.4$. Thus the initial state of the
adsorbate layer is still in the commensurate  $c(2\times 2)$ phase
and the diffusion dynamics in the bulk is exceedingly small. This is
the first hint that the boundary effects play a significant role in
the time evolution of the non-equilibrium profile. Beyond the value
of $r_0 \approx 1.56$, the refilling velocity rapidly takes off.

\begin{figure}[btp]
\includegraphics[bb= -0.5cm -0cm  9cm   7cm, width=7.5cm]{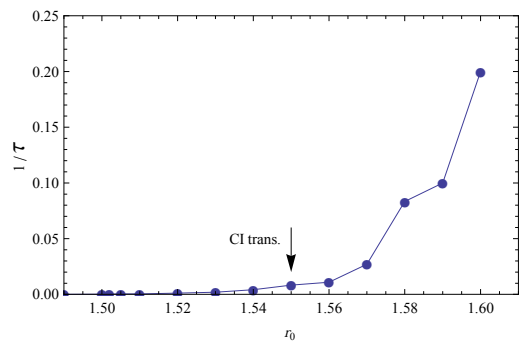}
\caption{Inverse refilling time $1/\tau$ (in units of $10^3$ MD time
steps) as a function of the LJ pair potential parameter $r_o$.  For
$L=50$, $T=0.4$ and coverage $\theta=0.5$.}
\end{figure}

To understand the role of the boundary effect qualitatively, we can
look at the known results of a semi-infinite  one-dimensional
Frenkel-Kontorova  chain with an open boundary at one end
\cite{Pokrov,frank}. Under a compressional strain, the last adatom
is pushed out towards the next maximum in the substrate potential
due to the lack of balancing force from the unoccupied site, with
the neighboring adatoms making small adjustments to reach
equilibrium. There is an activation barrier for the end atom to hop
over to the unoccupied side and thus creating a "light" domain wall
at the boundary. For increasing value of the parameter $r_0$, the
mismatch and the compressive stress increases and the end atom gets
pushed further out. Correspondingly, the activation barrier for the
"light" wall decreases and eventually this barrier can vanish
leading to the spontaneous generation of a "light" wall at the
boundary. At finite temperatures, the onset of rapid time evolution
of the non-equilibrium profile should correspond to the value of
$r_0$ at which the barrier is comparable to $k_B T$. Once generated,
this "light" wall can now move rapidly further into the chain under
the mechanism of biased diffusion due to the compressive stress. The
net effect is the motion of the boundary by one lattice spacing into
the open area. The whole process can then repeat itself leading to
the advancement of the profile and refilling of the initially empty
region. Obviously, the process will stop at some point if the chain
is finite. However, for a semi-infinite chain or one maintained at a
constant chemical potential at the far end, the unfilled region can
be filled at a steady rate via this mechanism. The creation of the
"light" wall at the boundary and its rapid propagation into the
interior is the key ingredient in understanding the anomalous
evolution of the nonequilibrium profile as shown in our numerical
study. Note that at the onset of this rapid anomalous transport,
normal bulk diffusion rate of domain walls and defects are
negligible in the interior of the adsorbate still in the
commensurate state. However, the "light" wall created at the
boundary can  diffuse inwards because the unbalanced boundary stress
essentially provides an inward driving force leading to a biased
diffusion. For the same reason, the profile is sharp because the
dynamics of its evolution is governed by a biased diffusion
mechanism rather than the normal diffusive behavior as observed in
typical Boltzman-Matano systems \cite{Loburets}. The activation
barrier for the "light" wall has been studied quantitatively in the
1D Frenkel-Kontorova model in the continuum limit
\cite{Pokrov,frank} but its detailed dependence on the mismatch  for
a fully discrete 2D model as presented in this study is yet unknown.

We have also studied the time evolution dynamics as a function of
the strength of the substrate potential as shown in Fig. 5. As the
substrate potential gets weaker, the rate of time evolution of the
profile increases. This can be interpreted as due to two factors.
The first is that the activation barrier of the "light" wall at the
boundary decreases with decreasing strength of the substrate
potential. The second factor is that the biased diffusion rate of
the "light" wall towards the interior is also enhanced through the
reduction of the substrate potential.

\begin{figure}[btp]
\includegraphics[bb= -0.5cm -0cm  9cm   6cm, width=7.5cm]{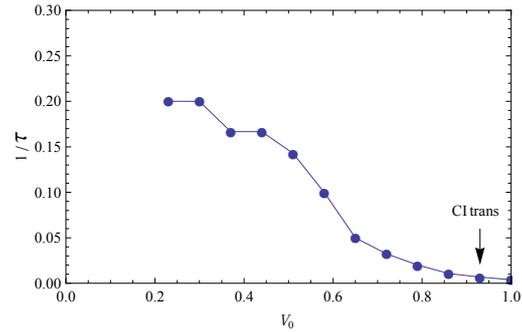}
\caption{Inverse refiling time $1/\tau$ (in units of $10^4$ MD time
steps) as a function of the strength of the substrate potential
$V_0$. For $L=50$, $r_o = 1.53$, $T=0.4$ and coverage $\theta=0.5$.}
\end{figure}

\subsection{ Non-equilibrium profile evolution as a function of coverage }

While the variation of the LJ potential parameter $r_0$ and hence
the lattice mismatch nicely illustrates the role that the boundary
effects play in determining the time evolution of the
non-equilibrium profile, experimentally variation of the lattice
mismatch is hard to realize. Instead, the easy parameter to vary in
a real system such as Pb/Si(111) is the coverage \cite{Man,conf}. In
this section, we study the effect of varying the coverage beyond the
ideal commensurate coverage of $\theta=1/2$ on the refilling
dynamics. Shown in the Figs. 6 and 7 are the results of the
refilling dynamics as a function of the coverage for a fixed value
of potential parameter $r_0=1.53$.

\begin{figure}[btp]
\includegraphics[bb= -0.5cm -0cm  10cm   11cm, width=7.5cm]{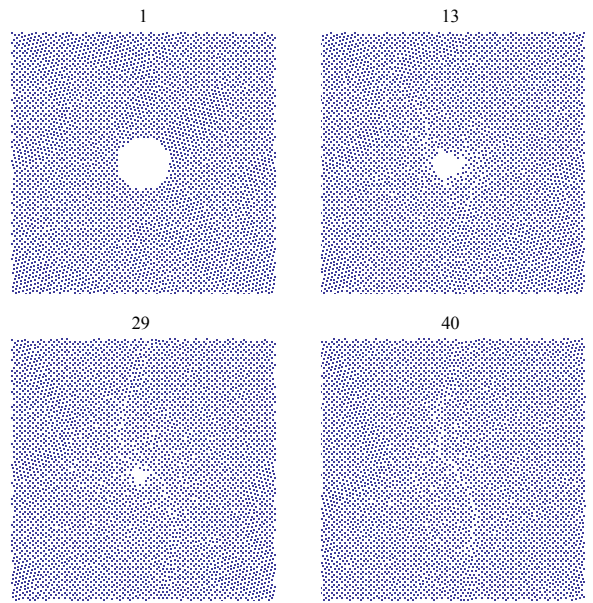}
\caption{Snapshots of the layer at different times during refilling
of the initial hole. For $L=100$, $r_o = 1.53$, $T=0.4$ and coverage
$\theta=0.508$.
The labels correspond to times in units of $3 \times 10^4$ MD time
steps.}
\end{figure}

\begin{figure}[btp]
\includegraphics[bb= -0.5cm -0cm  9cm  7cm, width=7.5cm]{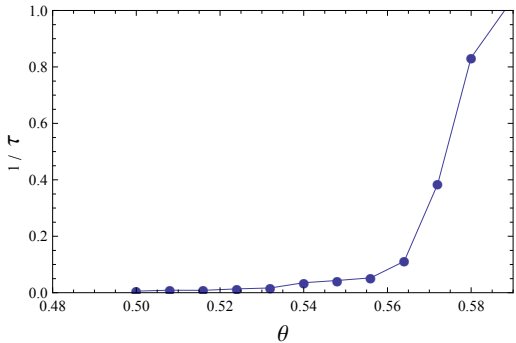}
\caption{Inverse refiling time $1/\tau$ (in units of $10^4$ MD time
steps) as a function of coverage $\theta$ for $L=100$, $r_o = 1.53$
at $T=0.4$.}
\end{figure}

As shown in Fig. 7, there is a critical coverage of $\theta \approx
0.5$ at T=0.4, below which the refilling dynamics is too slow to
measure in our study. Beyond a value of $\theta \approx 0.54$ the
refilling dynamics rapidly take off. Qualitatively, the increase of
coverage beyond $\theta=1/2$ again increases the compressive strain
and facilitates the creation of a light wall at the boundary, as
discussed in the last section. However, in this case another factor
come into consideration in determining the refilling dynamics.
Beyond the half coverage $\theta =1/2$, the configuration in the
bulk is now incommensurate and there are either defects or domain
walls (heavy type) to accomodate the extra coverage. For a system
below the Aubry transition \cite{Mazo,fisher,aubry1}, these defects
are pinned and immobile. In this phase, the diffusion of the "light"
wall created at the boundary into the interior will be blocked by
these defects and the refilling process will stop. As the coverage
increases, the critical potential strength for the Aubry transition
decreases \cite{Mazo} and these defects and domain walls become
mobile. At this point, the refilling dynamics can proceed smoothly
just as discussed in the last section where only the value of the
mismatch is varied. Thus, when the coverage is increased beyond the
ideal commensurate value of $\theta=1/2$, the time evolution of the
profile is determined by the convolution of both the activation rate
of domain walls at the boundary and the diffusion rate of defects in
the interior of the adsorbate layer. We like to point out that our
use of the term "critical coverage" in the present context differ
somewhat from what was used in the empirical description of the data
for Pb/Si(111)\cite{Man,conf} where the critical coverage was used
to describe the value of the coverage where the rate crossover from
a region with an activation barrier increasing with decreasing
coverage to another region with weak coverage dependence. In
particular, the details of the rapid increase of the profile
evolution rate beyond the critical coverage has not been
investigated experimentally.

\section{Conclusion}

In this paper we have investigated the time evolution of
non-equilibrium profile in an adsorbate system near a commensurate
to incommensurate transition. In the commensurate $c(2\times 2)$
phase, the mobility and diffusive rate for the bulk system is
essentially zero. However if the absorbate system is under
compressive strain due to lattice mismatch with the substrate, then
the barrier for activation of a "light" domain wall is reduced at
the boundary separating the adsorbate from the unoccupied region.
For increasing strain energy due to either increased lattice
mismatch or increase of coverage beyond the ideal coverage for the
commensurate $c(2\times 2)$ phase, the barrier for this boundary
domain wall generation will be lowered to the point where it can be
easily activated thermally. The profile can then evolve in time
through the repeated activation of the boundary domain wall and its
biased diffusion into the interior of the adsorbate layer. This
mechanism is similar to the one proposed by Huang {\it et al.}
\cite{Wang}. The difference is that in our study the coverage
dependence of the activation barrier of the boundary domain walls
comes out naturally as a consequence of the microscopic model and is
not invoked as an additional assumption. The profile remains sharp
during its evolution because it is continuously being driven by the
unbalanced compressive stress at the boundary. When the stress at
the boundary is increased via increasing the coverage beyond the
ideal commensurate coverage instead of changing the lattice
mismatch, an extra ingredient is required besides the activation of
the boundary domain wall. In this case, the interior of the
adsorbate layer has intrinsic defects or domain walls to accommodate
the extra adatoms. Depending on the substrate potential and the
coverage, these defects can be either pinned or depinned. The
transition between these two phases correspond to the "Aubry
transition " well studied in the literature
\cite{Mazo,fisher,aubry1}. When the domain walls are "pinned", the
propagation of the boundary generated domain wall into the interior
will be blocked by the pinned domain walls and the time evolution of
the non-equilibrium profile will also stop. The anomalous fast
transport can only be achieved when the system is in the unpinned
side of the incommensurate phase. The transition to this side of the
"Aubry phase" requires the coverage to exceed a critical value,
leading to an extra requirement for the the rapid propagation of the
non-equilibrium profile besides the activation of the boundary
domain wall. This aspect also differs from the generalized
one-dimensional Frenkel-Kontorova model results of Huang {\it et
al.} \cite{Wang} where domain walls in the bulk is always mobile and
unpinned beyond the monolayer coverage. Although the present study
is for a specific model system with a relatively simple
commensurate-incommensurate phase transition, the generic features
of the profile evolution is applicable to a wide class of adsorbate
layer under compressive strain. The mechanism of anomalous mass
transport via activation and propagation of boundary domain walls
should play an important role in tribological systems.

\begin{acknowledgements}
We acknowledge  helpful discussions with M.C. Tringides, M.S. Altman
C.Z. Wang, T. Ala-Nissila and  C. Zdenek.  E.G. was supported by
Funda\c c\~ao de Amparo \`a Pesquisa do Estado de S\~ao Paulo -
FAPESP (Grant No. 07/08492-9) and in part by computer facilities
from Centro Nacional de Processamento de Alto Desempenho -
CENAPAD-SP.

\end{acknowledgements}



\end{document}